\documentclass[12pt]{article}
\usepackage{amsmath}
\usepackage{latexsym}
\usepackage{a4wide}
\usepackage{bbm}
\newcommand{\I}{\text{i}}
\newcommand{\E}{\text{e}}

\newcommand{\re}[1]{(\ref{#1})}
\renewcommand{\cite}[1]{${}^{#1}$}
\begin{document}
\abovedisplayskip17pt plus2pt minus4pt
\abovedisplayshortskip14pt plus2pt minus4pt
\belowdisplayskip17pt plus2pt minus4pt
\belowdisplayshortskip14pt plus2pt minus4pt
\title{The Coulomb Green's Function in Two Dimensions}
\author{Walter Dittrich\\
  Institut f\"ur theoretische Physik, Universit\"at T\"ubingen,\\
  72076 T\"ubingen, Germany}
\maketitle
\begin{abstract}
We consider the two-dimensional non-relativistic Coulomb problem with
the aid of the momentum space construction of the associated Green's
function. Our presentation has precursors in three dimensions. It is
mainly Schwinger's approach which we advocate and project onto two
dimensions.
\end{abstract}

\section{Introduction}
The quantum mechanical Coulomb problem plays a central role in atomic
physics. Its solution is commonly studied using Schr\"odinger's wave
equation, wherein the bound state spectrum as well as the scattering
states can be exhibited by employing either spherical or parabolic
coordinates. One should not forget, however, that is was Pauli who
gave the first solution of the hydrogen atom with the aid of the
Laplace-Runge-Lenz vector which makes the H-atom a dynamical symmetry
problem in group theory. Finally there is the powerful method of
Green's functions which contains all information about the system. For
our case it means that if one is able to present a closed-form
expression for the Coulomb Green's function, one can immediately
extract the energy spectrum as well as the wave functions. Hence we
have to solve the Green's function equation or the associated integral
equation for the Coulomb potential.

In the sequel it is our goal to help the reader to analyze the Coulomb
problem once again using the language of Green's functions, but this
time in two spatial dimensions. Admittedly the three dimensional
problem is the physically important one and technically somewhat more
complicated than the two-dimensional case; it seems to us, however,
that a pedagogical discussion of the two-dimensional Green's function
for the Coulomb problem should be of wide interest.

Our treatment will be reminiscent of Schwinger's paper\cite{1}. There
are, however, calculations based on the hydrogen wave function in
momentum space that date back to V.A. Fock\cite{2} and V.
Bargmann\cite{3}. The paper by B. Podolsky and L. Pauling\cite{4} is
yet another important contribution. Along the same line of thought are
the articles furnished by the authors in Ref. 5 and 6. Two more recent
pedagogically noteworthy articles were provided by B.R.
Holstein\cite{7} and G.S.  Adkins\cite{8}. The nice review article by
M. Lieber\cite{9} and his contribution in Ref. 10 is also worth
mentioning. In the beginning there stands, of course, Pauli's seminal
work on the hydrogen atom\cite{11}.

\section{The 2-D Hydrogen Atom in Momentum Space and its Projection
  onto the Fock Sphere in 3-D}
Since the problem has already been discussed in this Journal\cite{12}
we will merely list some of the well-known results following from the
existence of the conserved Laplace-Runge-Lenz vector. But we will also
remind the reader of Fock's and Bargmann's work in the context of the
simpler two-dimensional Coulomb problem and so assist the student in
understanding their contribution as well. 

In two dimensions, the Hamiltonian is given by
\begin{equation}
H=\frac{\mathbf{p}^2}{2m} -\frac{Ze^2}{r}, \quad
\mathbf{p}^2=p^2_x+p^2_y, \quad r=\sqrt{x^2+y^2}. \label{1}
\end{equation}
The angular momentum vector has only one component, $L=L_3$, and the
Runge-Lenz vector degenerates to a two-dimensional vector,
$\mathbf{A}=(A_1,A_2)$ . In 2-D one finds ($\hbar=1$):
\begin{equation}
\mathbf{L\times p+p\times L} =\I \mathbf{p}\qquad \text{(not $2\I
  \,\mathbf{p}$ as in 3-D)}, \label{2}
\end{equation}
so that
\begin{equation}
\mathbf{A}=\frac{\mathbf{r}}{r}+\frac{1}{mZe^2} \left(
  -\mathbf{p\times L} +\frac{1}{2}\I\, \mathbf{p} \right) \label{3}
\end{equation}
and
\begin{equation}
\mathbf{A\times A}=\I \frac{(-2H)}{mZ^2e^4} \mathbf{L}, 
\label{4}
\end{equation}
i.e., 
\begin{equation}
[A_1,A_2]=\I \frac{(-2H)}{mZ^2e^4}\, L. 
\label{5}
\end{equation}

We are interested in the bound state spectrum $(-H>0)$ with the energy
values 
\begin{equation}
E'=-\frac{mZ^2e^4}{2\left(l+\frac{1}{2} \right)^2}, \quad
l=0,1,2,\dots, \quad l+\frac{1}{2}=:\nu, \label{6}
\end{equation}
so that we obtain for energy eigenstates according to Eq. \re{4}
\begin{equation}
\nu\mathbf{A\times}\nu\mathbf{A}=\I\, \mathbf{L}. \label{7}
\end{equation}
Recall that in 2-D the vector product is a pure number:
\begin{displaymath}
\mathbf{a\times b}=a_1b_2-a_2b_1=\#.
\end{displaymath}
Now it is useful to eliminate $\frac{1}{r}$ in favor of $\mathbf{p}^2$
and $H$:
\begin{equation}
\frac{1}{r}=\frac{1}{Ze^2} \left( \frac{\mathbf{p}^2}{2m} -H
\right)=\frac{1}{mZe^2} \left( \frac{\mathbf{p}^2}{2}-mH
\right). \label{8}
\end{equation}
Then Eq. \re{3} can be rewritten as
\begin{eqnarray}
mZe^2\mathbf{A}&=&\mathbf{r} \frac{mZe^2}{r}-
    \left(\underbrace{\mathbf{p\times L}}_{}-\frac{1}{2}\I\, \mathbf{p}
    \right) \nonumber\\
&&\qquad\qquad\quad =-\mathbf{p\times (p\times r)=-p\,\, p\cdot r+p^2
    \,r} \nonumber\\
&=&-\mathbf{r} \,mH+\frac{1}{2} \mathbf{r\, p^2+p\,\, p\cdot r -p^2\,r}
    +\frac{1}{2} \I\, \mathbf{p}.\label{9}
\end{eqnarray}
When acting on energy eigenstates one may write
\begin{equation}
H=E=-\frac{mZ^2e^4}{2\nu^2}=-\frac{\left(\frac{Z}{a_0}\right)^2}{2m}
\frac{1}{\nu^2}, \quad a_0=\frac{\hbar^2}{me^2}. \label{10}
\end{equation}
Introducing the effective momentum
\begin{equation}
p_0=\frac{Z}{a_0} \frac{1}{\nu}, \label{11}
\end{equation}
we can replace Eq. \re{10} by
\begin{equation}
-E=\frac{p_0^2}{2m}, \label{12}
\end{equation}
which yields for $mH$ in Eq. \re{9}: $-mH=\frac{p^2_0}{2}$. We also
rewrite in Eq. \re{9}
\begin{displaymath}
\frac{1}{2} \mathbf{r}\, \mathbf{p}^2=\frac{1}{2}\mathbf{p^2\, r}
+\frac{1}{2} [\mathbf{r,p^2}]=\frac{1}{2}\mathbf{p^2\, r} +\I\,
\mathbf{p}, 
\end{displaymath}
so that we obtain
\begin{equation}
mZe^2\mathbf{A}=\frac{1}{2}(p_0^2-p^2) \mathbf{r} +\mathbf{p\, p\cdot
  r} + \frac{3}{2}\I\, \mathbf{p}. \label{13}
\end{equation}
In 3-D one finds 2 instead of $\frac{3}{2}$ in the last term on the
right-hand side of Eq. \re{13}. 

At this stage we go to the momentum representation and write for any
state vector $|$ $\rangle$ the numerical value
\begin{equation}
\langle\mathbf{p}|\mathbf{A(r,p)}|\,\,\rangle=\mathbf{A} \left(\I
  \frac{\partial}{\partial \mathbf{p}}, \mathbf{p}\right)
  \langle\mathbf{p} |\,\,\rangle\equiv
  \mathbf{A}\,\psi(\mathbf{p}). \label{14}
\end{equation}
Consequently, the operator statement \re{13} turns into a differential
equation:
\begin{equation}
mZe^2\, \mathbf{A}\, \psi=\left[ \frac{1}{2}(p_0^2-p^2)\, \I
  \frac{\partial}{\partial \mathbf{p}} +\mathbf{p\, p\cdot} \I
  \frac{\partial}{\partial \mathbf{p}}+ \frac{3}{2} \I\, \mathbf{p}
  \right] 1\, \psi. \label{15}
\end{equation}
We now put $1=\frac{1}{(p_0^2+p^2)^{3/2}} {(p_0^2+p^2)^{3/2}}$ and
pull the denominator all the way to the left, past the differential
operator. Here we employ the formula
\begin{displaymath}
\frac{\partial}{\partial \lambda} f=f\frac{\partial}{\partial \lambda}
+\left(\frac{\partial f}{\partial \lambda} \right)
=f\frac{\partial}{\partial \lambda} +f \frac{\frac{\partial
    f}{\partial \lambda}}{f} =f\left( \frac{\partial}{\partial
    \lambda} +\frac{\partial}{\partial \lambda} \log f \right), 
\end{displaymath}
and upon using $f=\frac{1}{(p_0^2+p^2)^{3/2}}$ so that
$\frac{\partial}{\partial \mathbf{p}} \log \frac{1}{(p_0^2+p^2)^{3/2}}
=-3 \frac{\mathbf{p}}{(p_0^2+p^2)}$, we obtain for Eq. \re{15}:
\begin{equation}
p_0\nu\, \mathbf{A}\, \psi =\frac{1}{(p_0^2+p^2)^{3/2}} \I \left[
  \frac{1}{2} (p_0^2-p^2) \frac{\partial}{\partial \mathbf{p}}
  +\mathbf{p\, p\cdot}\frac{\partial}{\partial \mathbf{p}} \right]
  (p_0^2+ p^2)^{3/2}\, \psi. \label{16}
\end{equation}
On the left-hand side of this equation we made use of \re{11}:
$mZe^2=\frac{Z}{a_0} =\nu p_0$. 

The result \re{16} can be checked for the ground state $\mathbf{A}=0$,
which requires $(p_0^2+p^2)^{3/2}\, \psi_0=$const. or
$\psi_0(\mathbf{p})= \frac{\text{const.}}{(p_0^2+p^2)^{3/2}}$. This
is also the wave function found in Ref. 13, where the
two-dimensional analog of Fock's treatment is exhibited beautifully. 

In the sequel we will be interested in commutation relations
$[X,Y]=Z$. Say we have $\bar{X}=F^{-1}\,X\,F$, which is called a
similarity transformation, maintaining algebraic properties. We
encounter such a situation in Eq. \re{16} with
$F=(p_0^2+p^2)^{3/2}$. If we want simple commutation relations, we
must look in the middle, i.e., in the square brackets of \re{16}, as
the wings cannot be effective. To do so we consider a change of
variables to eliminate the $\mathbf{p\cdot} \frac{\partial}{\partial
  \mathbf{p}}$ term in \re{16}, which can be accomplished by
projecting the 2-D momentum space onto the 3-D sphere in the same way
that Fock did for the 3-D momentum space problem. Hence we introduce
the stereographic projection. 
\begin{equation}
\boldsymbol{\xi}=\frac{2p_0\mathbf{p}}{\lambda(p)}, \qquad \xi_0=
\frac{p_0^2-p^2}{\lambda(p)}, \qquad \lambda(p):=p_0^2+p^2, \label{17}
\end{equation}
where $\xi=(\xi_0,\boldsymbol{\xi})$ is a unit 3-vector, 
\begin{equation}
\boldsymbol{\xi}^2+\xi_0^2=1, \label{18}
\end{equation}
i.e., defines the unit sphere in a 3-D Euclidean space. In Eq. \re{16}
we need
\begin{displaymath}
\frac{\partial}{\partial  \mathbf{p}}=\frac{\partial
  \boldsymbol{\xi}}{\partial \mathbf{p}} \frac{\partial}{\partial
  \boldsymbol{\xi}}=2p_0 \left(\frac{1}{\lambda(p)}- \frac{2\mathbf{p\,
  p\cdot}}{\lambda(p)^2} \right) \frac{\partial}{\partial
  \boldsymbol{\xi}}.
\end{displaymath}
Then $\mathbf{p\, p\cdot} \frac{\partial}{\partial
  \mathbf{p}}=2p_0\, \mathbf{p} \frac{p_0^2-p^2}{\lambda(p)^2}\,
  \mathbf{p\cdot} \frac{\partial}{\partial
  \boldsymbol{\xi}}$. 

Using these results in \re{16} we obtain
\begin{equation}
\frac{1}{2} (p_0^2-p^2) \frac{\partial}{\partial
  \mathbf{p}}+ \mathbf{p\, p\cdot}\frac{\partial}{\partial
  \mathbf{p}}=p_0 \frac{p_0^2-p^2}{\lambda(p)} \frac{\partial}{\partial
  \boldsymbol{\xi}}=p_0\, \xi_0\, \frac{\partial}{\partial
  \boldsymbol{\xi}}. \label{19}
\end{equation}
Now Eq. \re{16} can be expressed in the form
\begin{eqnarray}
\nu\mathbf{A}\,\psi&=& \frac{1}{\lambda(p)^{3/2}} \I\,
  \xi_0\,\frac{\partial}{\partial \boldsymbol{\xi}}\, \lambda(p)^{3/2}\, \psi,
  \label{20}\\
\text{i.e.,} \qquad \nu\mathbf{A}&\to&\I\, \xi_0\, \frac{\partial}{\partial
  \boldsymbol{\xi}}, \label{21}
\end{eqnarray}

Because $\lambda(p)^{3/2}$ is a scalar we can also write $\mathbf{L}$
in terms of $\boldsymbol{\xi}$:
\begin{equation}
(\mathbf{L})_3\to \frac{1}{\I} \left(
  \mathbf{p\times}\frac{\partial}{\partial \mathbf{p}} \right)_3
  =\frac{1}{\I} \left(\boldsymbol{\xi\times}\frac{\partial}{\partial
  \boldsymbol{\xi}}\right)_3=\lambda(p)^{-3/2} \frac{1}{\I}
  \underbrace{\left(\boldsymbol{\xi\times}\frac{\partial}{\partial
  \boldsymbol{\xi}}\right)_3}_{=\xi_1\partial_2-\xi_2\partial_1}
  \lambda(p)^{3/2}. \label{22}
\end{equation}
So, except   for a similarity transformation in \re{20}, we have
\begin{eqnarray}
[\nu A_1,\nu A_2]&=&-\left[\xi_0 \frac{\partial}{\partial \xi_1},\xi_0
\frac{\partial}{\partial \xi_2} \right] \nonumber\\
&=&-\xi_0 \left[\frac{\partial}{\partial \xi_1},\xi_0\right]
\frac{\partial}{\partial \xi_2} -\xi_0 \left[\xi_0,\frac{\partial}{\partial
    \xi_2}\right] \frac{\partial}{\partial \xi_1}. \label{23}
\end{eqnarray}
From the constraint equation $\boldsymbol{\xi}^2+\xi_0^2=1$ we can use
$\frac{\partial}{\partial \boldsymbol{\xi}}
\xi_0=-\frac{\boldsymbol{\xi}}{\xi_0}$. Accordingly Eq. \re{23} reduces to
\begin{equation}
[\nu A_1,\nu A_2]=-\xi_0 \left( -\frac{\xi_1}{\xi_0} \right)
\frac{\partial}{\partial \xi_2} -\xi_0 \left( \frac{\xi_2}{\xi_0} \right)
\frac{\partial}{\partial \xi_1} =\xi_1 \partial_2 -\xi_2 \partial_1 =:\I\,
L_{12} \equiv \I\, L_3. \label{24}
\end{equation}
$\mathbf{A}$ in \re{21} and $\mathbf{L}$ on the right-hand side of
\re{24} look quite different. But it is possible to write them in the
same form. To do this we must get of the Fock sphere, which is our
unit sphere in 3-D space. Recall that, up until now, $\xi_0$ has not
been an independent variable: $\xi_0^2=1-\boldsymbol{\xi}^2$. Now let us think
of $\xi_0$ as being independent. Then the following obvious relation
exists between our former spatial derivative, where $\xi_0$ was
constrained, and a new derivative, where $\xi_0$ is now an independent
variable:
\begin{eqnarray}
\frac{\partial}{\partial \boldsymbol{\xi}}(\xi_0\,\,\text{constrained}) &\to&
  \frac{\partial}{\partial \boldsymbol{\xi}} +\frac{\partial \xi_0}{\partial
  \boldsymbol{\xi}} \frac{\partial}{\partial \xi_0} (\xi_0\,\,
  \text{independent}) \nonumber\\
&=&\frac{\partial}{\partial \boldsymbol{\xi}}-\frac{\boldsymbol{\xi}}{\xi_0}
  \frac{\partial}{\partial \xi_0} . \nonumber
\end{eqnarray}
Now we can write, instead of Eq. \re{21}, where $\xi_0$ is still a
dependent variable,
\begin{equation}
-\xi_0 \frac{1}{\I} \frac{\partial}{\partial \boldsymbol{\xi}} (\xi_0\,\,
\text{dep. variable}) \to \boldsymbol{\xi} \frac{1}{\I}
\frac{\partial}{\partial \xi_0} -\xi_0 \frac{1}{\I}
\frac{\partial}{\partial \boldsymbol{\xi}}(\xi_0\,\,
\text{indep. variable}). \label{25}
\end{equation}
Eq. \re{25} is just a rotation connecting the 0-axis with the $k$-th
axis ($k=1,2$). This, then, is the meaning of $\mathbf{A}$ as a
generator of rotation. Our whole algebra becomes evident when we write
\begin{eqnarray}
L_3&=&L_{12}, \quad \nu A_1=:L_{20}, \quad \nu A_2=:L_{01},
  \label{26}\\
\text{then}\qquad L_{ab}&=&\xi_a\frac{1}{\I} \frac{\partial}{\partial
  \xi_b} -\xi_b \frac{1}{\I} \frac{\partial}{\partial \xi_a}, \quad
  a,b=0,1,2, \label{27}
\end{eqnarray}
and a direct calculation yields
\begin{equation}
\frac{1}{\I} [L_{ab}, L_{cd}]=\delta_{ad}L_{cd} -\delta_{bd}
L_{ca}. \label{28} 
\end{equation}
So we have found
\begin{equation}
\nu\, \mathbf{A}\, \psi(p)=\frac{1}{(p_0^2+p^2)^{3/2}}
\mathbf{M}\,(p_0^2+p^2)^{3/2}\, \psi(p), \label{29}
\end{equation}
where $\mathbf{M}$ is the differential operator
\begin{equation}
\mathbf{M} =\boldsymbol{\xi}\frac{1}{\I} \frac{\partial}{\partial \xi_0} -\xi_0
\frac{1}{\I} \frac{\partial}{\partial \boldsymbol{\xi}}, \label{30}
\end{equation}
where $\boldsymbol{\xi}$ and $\xi_0$ are given by Eq. \re{17}. If we then put
\begin{equation}
M_1=\nu\, A_1=L_{20}, \quad M_2=\nu\, A_2=L_{01}, \quad L_3=L_{12},
\label{31}
\end{equation}
or $L:=(\mathbf{M}, L_3)$, we obtain the O(3) algebra
\begin{eqnarray}
\frac{1}{\I} (L\times L)&=&L, \label{32}\\
L^2=L_{12}^2+L_{20}^2+L_{01}^2&\equiv&L_3+ \mathbf{M}^2=\sum_{a,b}
L_{ab}^2. \label{33}
\end{eqnarray}
We know the eigenvalues of $L^2$ with $L$ satisfying \re{32}:
\begin{equation}
(L^2)'=(\hbar^2)\, l(l+1), \qquad l=0,1,2\dots\, .\label{34}
\end{equation}
The eigenfunctions are, of course, the spherical harmonics
$Y_{lm}(\Omega)$. So we obtain
\begin{equation}
L^2\, Y_{lm}=l(l+1)\, Y_{lm}, \label{35}
\end{equation}
and the quantum number $m$ can take all the integer values from $-l$
to $l$, so that the degeneracy of the energy state is $2l+1$.\bigskip

Finally we want to demonstrate that Pauli's treatment of the H-Atom
leads directly to the method developed by Fock\cite{2}. We hereby take
advantage of Bargmann's work\cite{3}, which we adopt for two spatial
dimensions. 

Consider the following calculations in a Euclidean space of
dimensionality D, in particular, D=3. Can we derive the result
\re{35}, using the 3-D angular momentum directly?

We found already that
\begin{displaymath}
L_{ab}=\xi_a\frac{1}{\I} \frac{\partial}{\partial
  \xi_b} -\xi_b \frac{1}{\I} \frac{\partial}{\partial \xi_a}.
\end{displaymath}
Squaring this expression, we obtain
\begin{equation}
\frac{1}{2} \sum_{a,b}L_{ab}^2=-\frac{1}{2}\sum_{a,b} (\xi_a \partial_b
-\xi_b \partial_a)^2 =-\sum_{a,b} (\xi_a\partial_b\xi_a\partial_b
-\xi_a\partial_b\xi_b\partial_a). \label{36}
\end{equation}
Let us rewrite this equation in terms of
\begin{displaymath}
\sum_{a}\xi_a^2=\xi^2, \quad\sum_a\partial_a^2=\partial^2, \quad \sum_a
\xi_a\partial_a =\xi\cdot\partial. 
\end{displaymath}
Notice $\partial_b \xi_a=\xi_a\partial_b+\delta_{ba}$, $\partial_b
\xi_b=\xi_b \partial_b +\delta_{bb}$. Then
\begin{eqnarray}
\frac{1}{2} \sum_{a,b}L_{ab}^2&=&-\sum_{a,b}(\xi_a\xi_a\partial_b
\partial_b+\delta_{ba} \xi_a\partial_b -\xi_b
\underbrace{\xi_a\partial_b}_{=\partial_b\xi_a-\delta_{ab}} \partial_a
-\delta_{bb} \xi_a \partial_a) \nonumber\\
&=&-\xi^2\partial^2 +(\xi\cdot \partial)^2+(D-2)
\xi\cdot\partial. \label{37}
\end{eqnarray}
We now want to find eigenvalue solutions for this differential
operator. Let $f$ be {\em a} solution with $\partial^2f=0$ with $f$
homogeneous in $x$ to some degree: $(\xi\cdot\partial) f=d\,f,\quad
d=0,1,2,\dots\,$. Then consider the special case D=3:
\begin{displaymath}
(\xi\cdot\partial)\,S(\xi)=l\,S(\xi), \qquad \partial^2 S=0. 
\end{displaymath}
This choice reduces Eq. \re{37} to
\begin{displaymath}
\frac{1}{2} \sum_{a,b}L_{ab}^2\, S(\xi)=\bigl[ l^2+(3-2)l\bigr]\,
S(\xi)=l(l+1)\, S(\xi), 
\end{displaymath}
where now $S(\xi)$ are the well-known spherical harmonics, and so indeed
we come back to Eq. \re{35}. Thus we have solved our eigenvalue
problem, 
\begin{displaymath}
\frac{1}{2} \sum_{a,b}L_{ab}^2\, S(\xi)=\lambda\, S(\xi), 
\end{displaymath}
with
\begin{displaymath}
\lambda=l(l+1)\, (\hbar^2), \qquad S(\xi)=Y_{lm}(\Omega). 
\end{displaymath}

\section{The 2-D Green's Function of the H-Atom on Momentum Space}

We begin the discussion of the 2-D hydrogen atom with the Green's
function equation in momentum space:
\begin{equation}
\langle\mathbf{p}|\left(E-H_0+\frac{Ze^2}{r}\right)G|\mathbf{p}'\rangle
=\langle\mathbf{p}|\mathbf{p}'\rangle. \label{38}
\end{equation}
Here we recall Eq. \re{1}. $H_0$ is the Hamiltonian for the free
particle: $H_0=\frac{\mathbf{p}^2}{2m}$. 

Obviously we need
\begin{equation}
\langle\mathbf{p}|\left(\frac{1}{r}\,G\right)|\mathbf{p}'\rangle =\int
d^2p'' \langle\mathbf{p}|\left(\frac{1}{r}\right)
|\mathbf{p}''\rangle\langle\mathbf{p}''
|G|\mathbf{p}'\rangle. \label{39} 
\end{equation}
One verifies directly\cite{13} that
\begin{equation}
  \langle\mathbf{p}|\left(\frac{1}{r}\right)|\mathbf{p}''\rangle
  =\frac{1}{2\pi |\mathbf{p-p''}|}, \label{40}
\end{equation}
so that
\begin{equation}
\langle\mathbf{p}|\left(\frac{1}{r}\,G\right) |\mathbf{p}'\rangle =
\frac{1}{2\pi} \int d^2p'' \frac{1}{|\mathbf{p-p''}|}\,
G(\mathbf{p'',p'}), \label{41} 
\end{equation}
and from Eq. \re{38}:
\begin{equation}
\left(E-\frac{\mathbf{p}^2}{2m} \right) G(\mathbf{p,p'}) +
\frac{Ze^2}{2\pi} \int d^2p'' \frac{1}{|\mathbf{p-p''}|}\,
G(\mathbf{p'',p'})=\delta^2(\mathbf{p-p'}). \label{42}
\end{equation}
This is our fundamental Green's function equation which we want to
solve, assuming
\begin{equation}
E=-\frac{p_0^2}{2m}, \label{43}
\end{equation}
i.e., we restrict ourselves for the time being to $E<0$ and define
$p_0=\sqrt{-2mE}$. 

At this stage we introduce the Fock-sphere once again and set the 2-D
momentum space into one-to-one correspondence to the surface of the
unit sphere in 3-D:
\begin{eqnarray}
\xi_1&\equiv&\frac{x}{p_0}= \sin\theta\,
  \cos\phi=\frac{2p_0p_x}{\lambda(p)}, \qquad \lambda(p)=p_0^2+p^2
  \label{44}\\
\xi_2&\equiv&\frac{y}{p_0}= \sin\theta\,
  \sin\phi=\frac{2p_0p_y}{\lambda(p)}\label{45}\\
\xi_0&\equiv&\frac{z}{p_0}=\cos\theta=\frac{p_0^2-p^2}{\lambda(p)},
  \label{46}
\end{eqnarray}
where
\begin{equation}
\xi_0^2+\boldsymbol{\xi}^2=1=\left(\frac{z}{p_0}\right)^2
+\left(\frac{x}{p_0}\right)^2 +\left(\frac{y}{p_0}\right)^2.\label{47}
\end{equation} 
The area element on the unit sphere is
\begin{displaymath}
d\Omega=\sin\theta\, d\theta d\phi=-d(\cos\theta)\, d\phi, 
\end{displaymath}
and upon using Eq. \re{46}, 
\begin{displaymath}
\frac{d\cos\theta}{dp}=-\frac{(2p_0)^2p}{\lambda^2},
\end{displaymath}
we obtain 
\begin{equation}
d\Omega=\left( \frac{2p_0}{\lambda}\right)^2 p\,dp\,d\phi=
\left( \frac{2p_0}{\lambda}\right)^2 d^2p. \label{48}
\end{equation}
One can also write
\begin{eqnarray}
d\Omega&=&2d^3\xi\, \delta(\xi^2-1)\nonumber\\
&\equiv&2d^2\xi\, d\xi_0\,
  \delta\bigl[\xi_0^2-(1-\boldsymbol{\xi}^2)\bigr] \nonumber\\
&=&2d^2\xi\frac{d\xi_0^2}{2|\xi_0|}\, \delta\bigl[\xi_0^2-
  (1-\boldsymbol{\xi}^2)\bigr] =\frac{d^2\xi}{|\xi_0|},
  \qquad\xi_0=\sqrt{1-\boldsymbol{\xi}^2}=\text{Eq. \re{46}}. \nonumber
\end{eqnarray}
It is easy to check that indeed $\int d\Omega=4\pi$. The delta
function connecting two points on the unit sphere is, according to
Eq. \re{48}, 
\begin{equation}
\delta(\Omega-\Omega')=\left(\frac{\lambda}{2p_0}\right)^2\,
\delta(\mathbf{p-p'}), \label{49}
\end{equation}
and the distance squared between two points $\xi,\xi'$ on the Fock
surface ($\xi\cdot\xi'=\cos\gamma$) is given by
\begin{eqnarray}
\left(2\sin \frac{\gamma}{2}\right)^2&=&(\xi-\xi')^2=(\xi_0-\xi_0')^2
  +(\boldsymbol{\xi-\xi'})^2 \nonumber\\
&=&\left(\frac{x-x'}{p_0}\right)^2 +\left(\frac{y-y'}{p_0}\right)^2
  +\left(\frac{z-z'}{p_0}\right)^2 \stackrel{\re{44}-\re{46}}{=}
  \frac{4p_0^2}{\lambda(p)\lambda(p')} (\mathbf{p-p'})^2. \label{50}
\end{eqnarray}
Then, if we define
\begin{equation}
G(\mathbf{p,p'})=-8mp_0^2 \frac{1}{\lambda(p)^{3/2}}\,
\Gamma(\Omega,\Omega')\, \frac{1}{\lambda(p')^{3/2}}, \label{51}
\end{equation}
we can rewrite Eq. \re{42} in the form
\begin{equation}
\Gamma(\Omega,\Omega')-\frac{Ze^2}{2\pi} \frac{m}{p_0} \int
d\Omega''\, \frac{1}{|\xi-\xi''|} \, \Gamma(\Omega'',\Omega')
=\delta(\Omega-\Omega') . \label{52}
\end{equation}
Upon using the Green's function equation
\begin{equation}
-\partial^2 D(\xi-\xi')=\delta(\xi-\xi'), \label{53}
\end{equation}
where
\begin{equation}
D(\xi-\xi')=\frac{1}{4\pi} \frac{1}{|\xi-\xi'|}, \label{54}
\end{equation}
the surface integral equation \re{52} becomes
\begin{equation}
\Gamma(\Omega,\Omega')-2\nu \int d\Omega''\,D(\xi-\xi'') \,
\Gamma(\Omega'',\Omega') =\delta(\Omega-\Omega') , \label{55}
\end{equation}
with
\begin{equation}
\nu=\frac{Ze^2m}{p_0}.\label{56}
\end{equation}
Here it is useful to recall\cite{14}
\begin{equation}
\frac{1}{|\xi-\xi'|}\Biggr|_{|\xi|=1=|\xi'|}\left(=\frac{1}{2\sin
    \frac{\gamma}{2}} \right)=\sum_{l,m} \frac{2\pi}{l+\frac{1}{2}}\,
    Y_{lm}(\Omega) \, Y^\ast_{lm}(\Omega'). \label{57}
\end{equation}
Then the Green's function \re{54} is exhibited as
\begin{equation}
D(\xi-\xi')=\frac{1}{2} \sum_{l,m} \frac{1}{l+\frac{1}{2}}\,
    Y_{lm}(\Omega) \, Y^\ast_{lm}(\Omega'). \label{58}
\end{equation}
Also remember the completeness relation of the spherical harmonics:
\begin{equation}
\delta(\Omega-\Omega')=\sum_{l,m} Y_{lm}(\Omega) \,
Y^\ast_{lm}(\Omega'), \label{59}
\end{equation}
and the normalization 
\begin{equation}
\int d\Omega''\, Y^\ast_{lm}(\Omega'')\,Y_{l'm'}(\Omega'')=
\delta_{ll'}\, \delta_{mm'}. \label{60}
\end{equation}
With this information one can easily verify that \re{55} is solved by
\begin{equation}
\Gamma(\Omega,\Omega')=\sum_{l,m}
\frac{Y_{lm}(\Omega)Y^\ast_{lm}(\Omega')}{1-\frac{\nu}{l+\frac{1}{2}}}
. \label{61}
\end{equation}
The poles of Eq. \re{61} yield the energy eigenvalues:
\begin{displaymath}
1-\frac{\nu}{l+\frac{1}{2}}=0, \qquad \nu=\frac{Ze^2m}{p_0}
\end{displaymath}
or $\nu=l+\frac{1}{2}$:
\begin{displaymath}
\nu^2=\frac{m^2Z^2e^4}{p_0^2}=\frac{m^2Z^2e^4}{-2mE}
=\left(l+\frac{1}{2} \right)^2. 
\end{displaymath}
So we obtain once again $(\hbar=1)$: 
\begin{eqnarray}
E_l&=&-\frac{mZ^2e^4}{2\left(l+\frac{1}{2}\right)^2}, \qquad
l=0,1,2,\dots, \label{62}\\
\text{or}\qquad E_n&=&-\frac{mZ^2e^4}{2\left(n-\frac{1}{2}\right)^2},
\qquad n=1,2,\dots\, .\label{63}
\end{eqnarray}
The normalized wave function follows from our result Eq. \re{61}: 
\begin{equation}
\int d\Omega\, |Y|^2=1=\int \left(\frac{2p_0}{\lambda(p)}\right)^2\,
d^2p\, |Y|^2. \label{64}
\end{equation}
Furthermore, $\int d\Omega\, \xi_0\, |Y|^2=0$, since under $\xi_k\to
-\xi_k$, $k=0,1,2$, we have $Y\to (-1)^lY$: $|Y|^2\to |Y|^2$. Hence
\begin{equation}
0=\int\left(\frac{2p_0}{\lambda(p)}\right)^2\,
d^2p\,\frac{p_0^2-p^2}{\lambda(p)}\,  |Y|^2. \label{65}
\end{equation}
Adding Eqs. \re{64} and \re{65} we obtain
\begin{displaymath}
1=\int\left(\frac{2p_0}{\lambda(p)}\right)^2\,
d^2p\,\underbrace{\left[1+\frac{p_0^2-p^2}{\lambda}\right]}_{
  =\frac{2p_0^2}{\lambda}}\, |Y|^2=\int d^2p \, 8p_0^4
\frac{|Y|^2}{\lambda(p)^3} . 
\end{displaymath}
This result can be used to write for the normalized momentum wave
function: 
\begin{equation}
\psi_{lm}(\mathbf{p})=\frac{\sqrt{8}p_0^2}{(p_0^2+p^2)^{3/2}} \,
Y_{lm}(\Omega_p), \qquad \int d^2p\, |\psi|^2=1, \label{66}
\end{equation}
where $p_0=\frac{mZe^2}{l+\frac{1}{2}}$. 

We now want to write $\Gamma(\Omega,\Omega')$ in a form that will be
easy to continue analytically. To do this, we note the generating
function for the Legendre polynomials
\begin{eqnarray}
\frac{1}{\sqrt{1-2x\mu+\mu^2}}=\sum_{l=0}^\infty \mu^l\, P_l(x),
\qquad |\mu|<1 \nonumber\\
\text{and}\qquad P_l(\cos \gamma)=\frac{4\pi}{2l+1} \sum_m
Y_{lm}(\Omega)Y^\ast_{lm}(\Omega'), \nonumber
\end{eqnarray}
so that
\begin{displaymath}
\frac{1}{\sqrt{1-2\mu\cos\gamma+\mu^2}}=\sum_{l=0}^\infty \mu^l\,
P_l(\cos\gamma) =\sum_l\mu^l \frac{4\pi}{2l+1} \sum_m
Y_{lm}(\Omega)Y^\ast_{lm}(\Omega') . 
\end{displaymath}
Using $-2\cos\gamma=-2\xi\cdot\xi'=(\xi-\xi')^2-2$, $|\xi|=|\xi'|=1$,
we get:
\begin{displaymath}
1-2\mu\cos\gamma+\mu^2=1+\mu\bigl[(\xi-\xi')^2-2\bigr]+\mu^2=(1-\mu^2)
+\mu(\xi-\xi')^2.
\end{displaymath}
This allows us to write:
\begin{equation}
\frac{1}{\sqrt{(1-\mu^2)+\mu(\xi-\xi')^2}} =\sum_{l,m} \mu^l
\frac{4\pi}{2l+1} \, Y_{lm}(\Omega)Y^\ast_{lm}(\Omega'). \label{67}
\end{equation}
Note, incidentally, that for $\xi=\xi'$ we obtain
\begin{eqnarray}
\frac{1}{1-\mu} \underbrace{\int d\Omega}_{=4\pi}&=&4\pi \sum_{l,m}
\frac{\mu^l}{2l+1} \,\int d\Omega\, |Y_{lm}(\Omega)|^2, \nonumber\\
\text{or}\qquad\frac{1}{1-\mu}&=&\sum_l\frac{\mu^l}{2l+1} \,\int
d\Omega  \sum_m|Y_{lm}(\Omega)|^2 =\sum_{l=0}^\infty
\frac{\mu^l}{2l+1} \, m(l), \nonumber
\end{eqnarray}
which again yields the multiplicity of the quantum number $l$:
\begin{eqnarray}
m(l)&=&2l+1, \qquad l=0,1,2,\dots \nonumber\\
\text{or}\qquad m(n)&=&2n-1, \qquad n=1,2,\dots\, .\nonumber
\end{eqnarray}
Now we return to our main result Eq. \re{61}. Use of the identity
\begin{displaymath}
\left(1-\frac{\nu}{l+\frac{1}{2}}\right)^{-1}=1+\frac{\nu}{l+\frac{1}{2}}
+\nu^2 \frac{1}{\left( l+\frac{1}{2}\right)\left(l+\frac{1}{2}-\nu
  \right)}  
\end{displaymath}
and the integral representation (valid for $\nu<1/2$): 
\begin{displaymath}
\frac{1}{(l+1)-(\nu+1/2)}=\int\limits_0^1d\mu\, \mu^{-(\nu+1/2)} \mu^l
\end{displaymath}
produces 
\begin{equation}
\Gamma(\Omega,\Omega')=\delta(\Omega-\Omega')+\frac{\nu}{2\pi}
\frac{1}{|\xi-\xi'|} +\frac{\nu^2}{2\pi} \int\limits_0^1 d\mu\,
\mu^{-(\nu+1/2)}
\frac{1}{\sqrt{(1-\mu^2)+\mu(\xi-\xi')^2}}. \label{68}
\end{equation}
Performing an integration by parts yields still another representation
for $\Gamma$:
\begin{equation}
\Gamma(\Omega,\Omega')=\delta(\Omega-\Omega')+\frac{\nu}{2\pi}
\int\limits_0^1 d\mu\, \mu^{-\nu} \frac{d}{d\mu} 
\frac{\mu^{\frac{1}{2}}}{\sqrt{(1-\mu^2)+\mu(\xi-\xi')^2}}. \label{69}
\end{equation}
Let us pause for a moment and look at the pole structure of
Eq. \re{68}:
\begin{eqnarray}
\int\limits_0^1 d\mu\,\mu^{-(\nu+1/2)}
\frac{1}{\sqrt{(1-\mu^2)+\mu(\xi-\xi')^2}} &=&
\frac{1}{\frac{1}{2}-\nu}
\frac{1}{\sqrt{(1-\mu^2)+\mu(\xi-\xi')^2}}\Biggr|_{\mu=0} \label{70}\\
&&-\int\limits_0^1 d\mu\, \frac{\mu^{1/2-\nu}-1}{\frac{1}{2}-\nu}\,
\frac{d}{d\mu}
\frac{1}{\sqrt{(1-\mu^2)+\mu(\xi-\xi')^2}}. \nonumber
\end{eqnarray}
Introducing Eq. \re{68} in Eq. \re{51} we obtain
\begin{eqnarray}
G(\mathbf{p,p'})=-8mp_0^2\frac{1}{\lambda(p)^{3/2}} \Biggl[&&\!\!\!\!
  \!\!
  \delta(\Omega-\Omega') +\frac{\nu}{2\pi} \frac{1}{|\xi-\xi'|}
  \label{71}\\
&&  +\frac{\nu^2}{2\pi} \int\limits_0^1 d\mu\, \mu^{-(\nu+1/2)}
  \frac{1}{\sqrt{(1-\mu^2)+\mu(\xi-\xi')^2}} \Biggr]
  \frac{1}{\lambda(p')^{3/2}} . \nonumber
\end{eqnarray}
Here, we consider only the $\mu$-integral term which yields, with the
aid of Eq. \re{70}: 
\begin{equation}
G(\mathbf{p,p'})=\dots -8mp_0^2\frac{1}{\lambda(p)^{3/2}}
\frac{\nu^2}{2\pi} \left[ \frac{1}{\frac{1}{2}-\nu}-\int\limits_0^1
  d\mu\, \frac{\mu^{1/2-\nu}-1}{\frac{1}{2}-\nu}\, \frac{d}{d\mu}
 \frac{1}{\sqrt{{\scriptstyle(1-\mu^2)+
       \mu(\xi-\xi')^2}}} \right]   
  \frac{1}{\lambda(p')^{3/2}} . \label{72}
\end{equation}
The pole contribution is obviously contained in 
\begin{displaymath}
G(\mathbf{p,p'})= -8mp_0^2\frac{1}{\lambda(p)^{3/2}}\,
\frac{\nu^2}{2\pi}\, \frac{1}{\frac{1}{2}-\nu}\,
\frac{1}{\lambda(p')^{3/2}} +\dots\, ,
\end{displaymath}
where $\nu=\frac{1}{2}$ corresponds to the ground state $n=1$:
$E_1=-2mZ^2e^4$. 

Recall $\frac{1}{2}=\nu=\frac{Ze^2m}{p_0}$: $p_0=2Ze^2m$,
$\nu^2=\frac{Z^2e^4m}{-2E}$, so that $\frac{1}{\frac{1}{2}-\nu}=
\frac{\frac{1}{2}+\nu}{\frac{1}{4}-\nu^2}$,
\begin{eqnarray}
\Longrightarrow\quad \frac{1}{\frac{1}{4}-\nu^2}&=&
\frac{4}{1-4\nu^2} =\frac{4}{1+\frac{2Z^2e^4m}{E}}
=\frac{4E}{E+2me^4Z^2} =\frac{4E}{E-E_1} \nonumber\\
&=&-\frac{2mZ^2e^4}{\nu^2} \frac{1}{E-E_1}, \nonumber\\
\text{and}\quad \frac{\nu^2}{2\pi} \frac{1}{\frac{1}{2}-\nu}
&=&\frac{1}{2\pi} \bigl( -2mZ^2e^4 \bigr)\, \frac{1}{E-E_1}. \label{73}
\end{eqnarray}
We need this result in Eq. \re{72}:
\begin{displaymath}
-8mp_0^2\frac{\nu^2}{2\pi} \frac{1}{\frac{1}{2}-\nu}
=\frac{2p_0^4}{\pi} \frac{1}{E-E_1}. 
\end{displaymath}
Hence we obtain:
\begin{equation}
G_{\text{E}}(\mathbf{p,p'})=\frac{1}{\lambda(p)^{3/2}}
\frac{2p_0^4}{\pi} \frac{1}{E-E_1} \frac{1}{\lambda(p')^{3/2}}+\dots\,
.\label{74}
\end{equation}
The remaining integral is defined for all $\nu$ such that Re
$\nu<\frac{3}{2}$. This process can be repeated as often as necessary
to isolate more poles and extend the acceptable region for $\nu$.

So far we have been interested in bound states. But from now on we
will be interested in scattering states. Hence we extend $\nu$
analytically to complex values, in particular to the imaginary
axis. So let us define
\begin{equation}
\eta=-\I\nu=\frac{mZe^2}{k}, \qquad k=\sqrt{2mE}\qquad
(E>0). \label{75}
\end{equation}
Again we go back to our fundamental equation \re{51} and use
Eq. \re{69} for $\Gamma(\Omega,\Omega')$:
\begin{eqnarray}
G(\mathbf{p,p'})&=&-8mp_0^2\frac{1}{\lambda(p)^{\frac{3}{2}}}\!
\left[\delta(\Omega-\Omega')+\frac{\nu}{2\pi} 
\int\limits_0^1 d\mu\, \mu^{-\nu} \frac{d}{d\mu} 
\frac{\mu^{\frac{1}{2}}}{\sqrt{(1-\mu^2)+\mu(\xi-\xi')^2}} \right]\!
\frac{1}{\lambda(p')^{\frac{3}{2}}} \nonumber\\
&=&-8mp_0^2\frac{1}{\lambda(p)^{3/2}} \left( \frac{\lambda(p)}{2p_0}
\right)^2 \, \delta(\mathbf{p-p'}) \,\frac{1}{\lambda(p')^{3/2}}+
\dots\label{76}\\
&=&-2m\, \frac{1}{\lambda(p)}\,  \delta(\mathbf{p-p'})+\dots \,
.\nonumber
\end{eqnarray}
Here we write $\lambda(p)=p_0^2+p^2=p^2-2mE=2m \left(
  \frac{\mathbf{p}^2}{2m} -E \right) =2m (T-E)$, so that
$G(\mathbf{p,p'})= \frac{1}{E-T}\, \delta(\mathbf{p-p'})+\dots$, with
$T=\frac{\mathbf{p}^2}{2m}$. This amounts to writing ($\nu=\I\eta$,
$p_0=-\I k$)
\begin{eqnarray}
G(\mathbf{p,p'})&=&\frac{\delta(\mathbf{p-p'})}{E-T}-8mp_0^2
  \frac{1}{\lambda(p)^{3/2}} \frac{mZe^2}{2\pi p_0} 
  \int\limits_0^1 d\mu\, \mu^{-\I\eta} \frac{d}{d\mu} 
  \frac{\mu^{\frac{1}{2}}}{\sqrt{(1-\mu^2)+\mu(\xi-\xi')^2}}
  \frac{1}{\lambda(p')^{3/2}} \nonumber\\
&=&\frac{\delta(\mathbf{p-p'})}{E-T} \label{77}\\
&&-\frac{Ze^2}{\pi}p_0 \frac{1}{E-T}
  \int\limits_0^1 d\mu\, \mu^{-\I\eta} \frac{d}{d\mu} 
  \frac{\mu^{\frac{1}{2}}}{\sqrt{[(1-\mu^2)+\mu(\xi-\xi')^2]
  \lambda(p)\lambda(p')}} \, \frac{1}{E-T'}.\nonumber
\end{eqnarray}
using relation \re{50}
\begin{displaymath}
(\xi-\xi')^2=\frac{4p_0^2}{\lambda(p)\lambda(p')}\, (\mathbf{p-p'})^2,
\end{displaymath}
the square root in \re{77} can also be rewritten as
\begin{equation}
2p_0\sqrt{(\mathbf{p-p'})^2\mu-\frac{m}{2E} (E-T)(E-T') (1-\mu)^2},
\label{78}
\end{equation}
so that we now have
\begin{eqnarray}
G(\mathbf{p,p'})&=&\frac{\delta(\mathbf{p-p'})}{E-T}\label{79}\\
&&-\frac{Ze^2}{2\pi}p_0 \frac{1}{E-T} \int\limits_0^1 d\mu\,
\mu^{-\I\eta} \frac{d}{d\mu}  \frac{\mu^{\frac{1}{2}}}{
  \sqrt{{\scriptstyle (\mathbf{p-p'})^2\mu-\frac{m}{2E} (E-T)(E-T')
      (1-\mu)^2}}} \frac{1}{E-T'}. \nonumber
\end{eqnarray}
Performing an integration by parts, it is easy to show that the
$\mu$-integral can be written as
\begin{equation}
\frac{1}{2p_0|\mathbf{p-p'}|} +\I\eta\int\limits_0^1 d\mu\,
\mu^{-(\I\eta +1/2)} \frac{1}{2p_0
  \sqrt{(\mathbf{p-p'})^2\mu-\frac{m}{2E} (E-T)(E-T') (1-\mu)^2}}
. \label{80}
\end{equation}
Because the scattering is characterized by
\begin{equation}
(E-T)\sim 0\sim (E-T'), \qquad (\mathbf{p-p'})^2>0, \label{81}
\end{equation}
we can replace the square root in \re{80} (read together with the
$\mu$-integral) by
\begin{equation}
\frac{1}{\sqrt{(\mathbf{p-p'})^2\mu-\frac{m}{2E} (E-T)(E-T')}}
  =\frac{1}{\I \sqrt{\frac{m}{2E}(E-T)(E-T')}} \,
  \frac{1}{\sqrt{1-\beta\mu}} \label{82}
\end{equation}
with
\begin{equation}
\beta:=\frac{(\mathbf{p-p'})^2}{\frac{m}{2E}(E-T)(E-T')}. \label{83}
\end{equation}
So the $\mu$-integral in Eq. \re{79} is given by
\begin{equation}
\frac{1}{2p_0|\mathbf{p-p'}|} +\frac{\eta}{2p_0} \frac{1}{
  \sqrt{\frac{m}{2E}(E-T)(E-T')}}  \int\limits_0^1 d\mu\, \mu^{-\I\eta
  -\frac{1}{2}}\, (1-\beta\mu)^{-\frac{1}{2}}. \label{84}
\end{equation}
In the limit of large $\beta$ the integral in \re{84} may be computed
with some formulas given in Ref. 15. As an intermediate result for our
Green's function we then obtain
\begin{equation}
G(\mathbf{p,p'})=\frac{\delta(\mathbf{p-p'})}{E-T}-G_0^{\text{C}}(\mathbf{p})
\left[ \frac{m}{(2\pi)^2\sqrt{\pi}}\, \frac{\E^{\I\eta \ln
      \frac{(\mathbf{p-p'})^2}{4p_0^2} }}{|\mathbf{p-p'}|} \,
  \frac{\Gamma \bigl( 1/2 -\I\eta\bigr)}{\Gamma(\I\eta)}\,
  \I^{-2\I\eta} \right] G_0^{\text{C}}(\mathbf{p'}) \label{85}
\end{equation}
where
\begin{equation}
G_0^{\text{C}}(\mathbf{p})=\frac{\sqrt{2\pi\I k}}{m} \,
\frac{\I^{(1+\I\eta)} \Gamma(1+\I\eta)}{E-T} \, \E^{ -\I\eta \ln
  \frac{E-T}{4E}}. \label{86}
\end{equation}
Incidentally, when we take the Fourier transform of this expression we
get (for large $r$):
\begin{equation}
\int d^2p\, \E^{\I\mathbf{p\cdot r}}\,
G_0^{\text{C}}(\mathbf{p})=\frac{1}{\sqrt{r}} \, \E^{\I(kr+\eta\ln
  2kr)}. \label{87}
\end{equation}
Since $(\mathbf{p-p'})^2=4p_0^2\sin^2 \frac{\phi}{2}$, where we
assumed $p^2=p'{}^2\simeq p_0^2=-2mE=-k^2$, $p_0=-\I k$, we have
$|\mathbf{p-p'}|=\sqrt{4p_0^2\sin^2\frac{\phi}{2}}$, so that the
square brackets in Eq. \re{85} take the value
\begin{equation}
\frac{m}{2\pi \sqrt{\I k}} \frac{1}{(2\pi)^{3/2}} \, \tilde{f}(\phi)\,
\I^{-2\I\eta}, \label{88}
\end{equation}
with 
\begin{equation}
\tilde{f}(\phi)=\frac{\E^{\I\eta\ln\sin^2 \frac{\phi}{2}}}{ \sqrt{2\I
    k\sin^2 \frac{\phi}{2}}}\, \frac{\Gamma\bigl( 1/2
    -\I\eta\bigr)}{\Gamma(\I\eta)} . \label{89}
\end{equation}
If we then choose the phase such that
\begin{displaymath}
\E^{\I\left[ \text{arg}\, \Gamma\bigl( 1/2
    -\I\eta\bigr)- \text{arg}\, \Gamma(\I\eta)\right]} =(\I k
    )^{\frac{1}{2}}\, \I^{2\I\eta}, 
\end{displaymath}
we finally obtain for the 2-D Coulomb Green's function:
\begin{equation}
G_{\text{E}}(\mathbf{p-p'})=\frac{\delta(\mathbf{p-p'})}{E-T}
-G_0^{\text{C}}(\mathbf{p}) \frac{m}{2\pi}
\frac{1}{(2\pi)^{\frac{3}{2}}} f(\phi)\, G_0^{\text{C}}(\mathbf{p'}),
\label{90} 
\end{equation}
where
\begin{equation}
f(\phi)=\frac{|\Gamma\bigl( 1/2 -\I\eta\bigr)|}{|\Gamma(\I\eta)|} 
\frac{\E^{\I\eta\ln\sin^2 \frac{\phi}{2}}}{ \sqrt{2\I
    k\sin^2 \frac{\phi}{2}}}. \label{91}
\end{equation}
This is the scattering amplitude that is needed to compute the
differential cross section
\begin{equation}
\sigma(\phi)=|f(\phi)|^2. \label{92}
\end{equation}
Using the formulas
\begin{displaymath}
|\Gamma(\I\eta)|^2=\frac{\pi}{\eta\sinh \eta\pi}, \qquad |\Gamma\bigl(
1/2-\I\eta)|^2=\frac{\pi}{\cosh \eta\pi}, 
\end{displaymath}
we get
\begin{equation}
\sigma(\phi)=\frac{\eta \tanh \eta\pi}{2k \sin^2 \frac{\phi}{2}},
  \label{93}
\end{equation}
and with $\eta=\frac{mZe^2}{k}$ we arrive at
\begin{equation}
\sigma(\phi)=\frac{mZe^2}{2k^2\sin^2 \frac{\phi}{2}}\, \tanh \frac{\pi
  mZe^2}{k}. \label{94}
\end{equation}
In the high energy limit ($k\to\infty$, $\eta\to 0$) we then obtain
\begin{eqnarray}
\sigma(\phi)&=&\frac{mZe^2}{2k^2\sin^2 \frac{\phi}{2}}\,  \frac{\pi
  mZe^2}{k}\qquad\qquad k=\frac{mv}{\hbar} \nonumber\\
&=&\frac{\pi(Ze^2)^2}{2\hbar mv^3 \sin^2 \frac{\phi}{2}}. \label{95}
\end{eqnarray}
This agrees with the Born approximation. Our result was also found in
Ref. 16.

\section{Conclusion}
In this paper we have studied the quantum mechanical Coulomb problem
in two spatial dimensions. Although it is true that the
three-dimensional analogue is the more important -- since physical --
one, it seems to us that the two-dimensional model helps immensely to
understand the mathematical aspect of the real three-dimensional
case. Spherical harmonics on the two-sphere are certainly more familiar
than the ones on the three-sphere. Following the strategy initiated by
Fock, Pauli and Schwinger, we were able to solve the two-dimensional
Coulomb problem analytically, i.e., we presented the exact Green's
function for the two-dimensional hydrogen atom. Exact formulas were
then given for both the discrete and the continuous parts of the
spectrum. We hope that by studying the present paper the reader will
not have any problem reproducing Schwinger's superb paper\cite{1} on
the same subject.
\vspace{2cm}

\noindent
ACKNOWLEDGEMENT\medskip

\noindent
The author is grateful to H. Gies for carefully reading the
manuscript.


\begin{thebibliography}{99}
\bibitem{1} J. Schwinger, {\em Coulomb Green's function},
  J. Math. Phys. {\bf 5}, 1606-1608 (1964).
\bibitem{2} V.A.Fock, {\em Zur Theorie des Wasserstoffatoms},
  Zeit. Phys. {\bf 98}, 145-154 (1935).
\bibitem{3} V. Bargmann, Zeit. Phys. {\bf 99}, 576-582 (1936).
\bibitem{4} B. Podolsky and L. Pauling, {\em The momentum distribution
  in hydrogen atoms}, Phys. Rev. {\bf 34}, 109-116 (1929).
\bibitem{klein} J.J. Klein, {\em Eigenfunctions of the Hydrogen Atom
  in Momentum Space}, Am. J. Phys. {\bf 34}, 1039-1042 (1966).
\bibitem{lange} O.L. de Lange and R.E. Raab, {\em An operator solution
  for the hydrogen atom with application to the momentum
  representation}, Am. J. Phys. {\bf 55}, 913-917 (1987).
\bibitem{5} B.R. Holstein, {\em Quantum Mechanics in Momentum Space:
  The Coulomb System}, Am. J. Phys. {\bf 63}, 710-716 (1995).
\bibitem{6} G.S. Adkins, {\em Derivation of the Schr\"odinger-Coulomb
  Green's function from the scattering expansion}, Nuov. Cim. B {\bf
  97}, 99-107 (1987).
\bibitem{7} M. Lieber, {\em The Coulomb Green's function}, in {\em
  Relativity, Quantum Electrodynamics and Weak Interaction Effects in
  Atoms}, edited by W. Johnson, P. Mohr, and J. Sucher, AIP Conference
  Proceedings No. 189 (AIP, New York, 1989), 445-459 (1989).
\bibitem{8} M. Lieber, {\em O(4) symmetry of the hydrogen atom and the
  Lamb shift}, Phys. Rev. {\bf 174}, 2037-2054 (1968).
\bibitem{9} W. Pauli, Jr., {\em \"Uber das Wasserstoffspektrum vom
  Standpunkt der Neuen Quantenmechanik}, Zeit. Phys. {\bf 36}, 336-363
  (1926).
\bibitem{10} X.L. Yang, M. Lieber, and F.T. Chan, {\em The Runge-Lenz
  vector for the two-dimensional hydrogen atom}, Am. J. Phys. {\bf
  59}, 231-232 (1991).
\bibitem{11} T.-I. Shibuya and C.E. Wulfman, {\em The Kepler problem
  in two-dimensional momentum space}, Am. J. Phys. {\bf 33}, 570-574
  (1965).
\bibitem{jack} J.D. Jackson, {\em Classical Electrodynamics}, Second
  Edition, (John Wiley\&Sons, Inc., New York, 1975), p.102, (3.70)
  or\\
  J. Schwinger, L.L. DeRaad, Jr., K.A. Milton, W.-y. Tsai, {\em
  Classical Electrodynamics}, (Perseus Books, Reading, Massachusetts,
  1998), p.246, (21.26).
\bibitem{12}  I.S. Gradshteyn and I.M. Ryzhik, {\em Tables of
  Integrals, Series and Products}, (Academic Press, New York, 1965),
  Refs. 3.197.3, 9.132.2 (1965).
\bibitem{13} G. Barton, {\em Rutherford scattering in two dimensions},
  Am. J. Phys. {\bf 51}, 420-422 (1983);\\
  Q.-g. Lin, {\em Scattering by a Coulomb field in two
  dimensions}, Am. J. Phys. {\bf 65}, 1007-1009 (1997).
\end{thebibliography}
\end{document}